\documentclass[twocolumn,prl, showpacs]{revtex4}
\usepackage{graphics,graphicx}
\begin{document}
\title{Nonconventional Spin Glass Transition in a Chemically Ordered Pyrochlore}
\author{D.K.~Singh$^{1,2,3}$}
\author{Y.S.~Lee$^{1}$}
\affiliation{$^{1}$Department of Physics, Massachusetts Institute of Technology, Cambridge, MA 02139}
\affiliation{$^{2}$NIST Center for Neutron Research, Gaithersburg, MD 20874}
\affiliation{$^{3}$Department of Materials Science and Engineering, University of Maryland, College Park, MD 20742}

\begin{abstract}

We report on the study of unusual spin glass properties in the geometrically frustrated pyrochlore Tb$_{2}$Mo$_{2}$O$_{7}$, $T$$_{g}$$\simeq$24 K. The analysis of the nonlinear part of dc and complex susceptibilities, near the glass transition regime, suggests the existence of a statistical distribution of relaxation times in short-range ordered ferromagnetic clusters. In addition, the magnetic spins are not sufficiently frozen below the glass transition temperature; apparently, responsible for the non-equilibrium scaling behavior of the static critical exponents of nonlinear susceptibilities. Our report is expected to shed new light in understanding the freezing properties of frustrated pyrochlores with short range ferromagnetic interactions.

\end{abstract}

 \maketitle

Geometrically frustrated magnets have attracted significant attention in recent years because of their intriguing physical and magnetic properties, that are often associated to the novel phenomena of spin liquid, spin ice or the spin glass transition at low temperatures.\cite{Joel,Ramirez2,Jason2} The spin glass state in geometrically frustrated compounds of stoichiometric composition, such as Tb$_{2}$Mo$_{2}$O$_{7}$, Y$_{2}$Mo$_{2}$O$_{7}$, Gd$_{3}$Ga$_{5}$O$_{12}$ (GGG) and SrCr$_{8}$Ga$_{4}$O$_{19}$ (SCGO),\cite{Gaulin,Gingras,Schiffer,Ramirez} is of particular interest. The absence of apparent chemical disorder in the underlying lattice does not fit congruently with a conventional understanding of the spin glass phenomenon, which requires the presence of chemical disorder to create a randomly frustrated system.\cite{Young} Recent research works on understanding the spin glass properties in geometrically frustrated compounds with antiferromagnetic ground state, such as Y$_{2}$Mo$_{2}$O$_{7}$, have suggested the presence of uncompensating magnetic interactions coupled with the random strains in the compound as the possible cause of this unusual behavior.\cite{Andreanov} However, the proposed theoretical interpretation does not explain the spin glass properties in geometrically frustrated compounds with short-range ferromagnetic order, for example Tb$_{2}$Mo$_{2}$O$_{7}$ ($T_g$ $\simeq$ 24 K).\cite{Martinez,Gaulin,Deepak,Ehler}  

Despite a lot of efforts, very little is known about the spin glass phenomenon in Tb$_{2}$Mo$_{2}$O$_{7}$\cite{Jiang,Ehler,Deepak}. A thorough understanding of the spin glass phenomenon in a material requires the analysis of the thermodynamic properties near the glass transition. Here we report on the detailed investigation of the spin glass transition in single crystal Tb$_{2}$Mo$_{2}$O$_{7}$ using nonlinear part of dc and ac susceptibilities measurements. While the analysis of the nonlinear dc susceptibilities allows us to understand the static properties near the spin glass transition, the ac susceptibilities provide important information about the dynamic properties in the spin glass state. The pyrochlore Tb$_{2}$Mo$_{2}$O$_{7}$ belongs to the $Fd\bar{3}m$ cubic space group, where both Tb$^{4+}$ and Mo$^{3+}$ sublattices form three dimensional interpenetrating network of corner sharing tetrahedra. Neutron scattering measurements on both powder and single crystal identified the ferromagnetically correlated short-range ordered clusters of Tb moments well below the spin glass transition $\simeq 1.5$~K.\cite{Greedan,Ehler,Deepak,Jiang} The static moment associated with the short-range order of Tb-ions at $T$ = 1.6 K was determined to be $\langle M_{Tb} \rangle \simeq 4.0(5)~\mu_B$, significantly smaller than that expected for a free Tb$^{3+}$ ion ($\sim9.5~\mu$$_{B}$).\cite{Deepak} We show that the spin glass transition in Tb$_{2}$Mo$_{2}$O$_{7}$ is a non-equilibrium phenomenon, which is driven by the statistical distribution of relaxation times in weakly interacting or independent magnetic clusters. In addition, spins are not sufficiently frozen even at the lowest measurement temperature (2 K), which suggests the presence of active spin dynamics in the spin glass state. Our findings are in direct contrast with the observation of a true equilibrium phase transition in an isostructural pyrochlore compound Y$_{2}$Mo$_{2}$O$_{7}$,\cite{Gingras} and can be attributed to the different nature of magnetic interactions in these compounds.

\begin{figure}
\centering
\includegraphics[width=12cm]{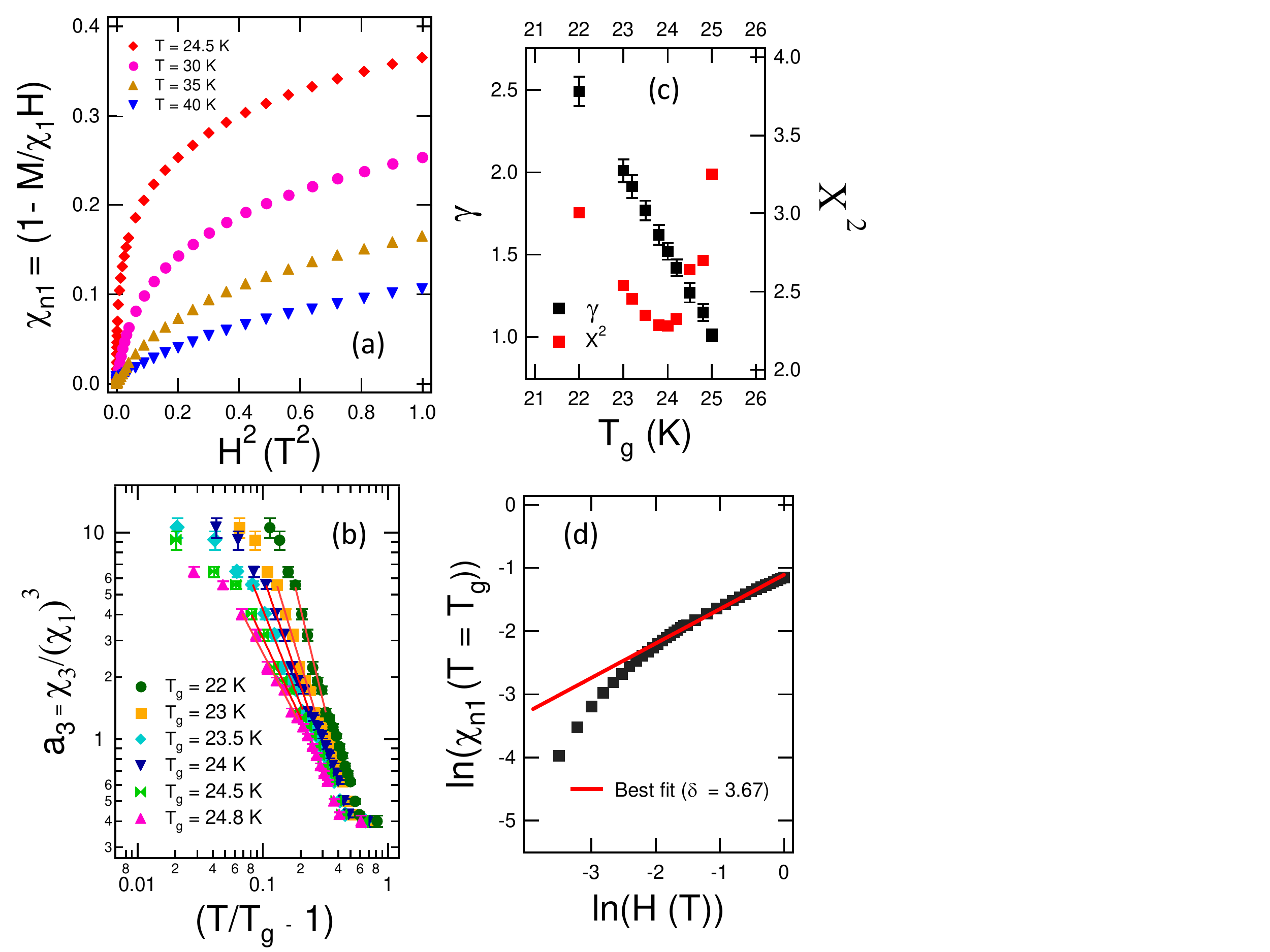} \vspace{-4mm}
\caption{(color online) (a) Net nonlinear susceptibilities at different temperatures, $\chi$$_{n1}$(T), as function of $H$$^{2}$. (b) The temperature dependencies of $a$$_{3}$ (= $\chi_3$/$\chi_1$$^{3}$) on log-log axes for few different selection of glass transition temperatures. Red curves are the best fits to a subset of the experimental data using Eq. (5) (see text for detail). (c) Extracted critical exponent $\gamma$ and the least square measure of the fitting parameter $X$$^{2}$ are plotted as a function of $T_g$. A reasonable good fit is obtained at $T$$_{g}$ $\simeq$ 24 K. (d) Plot of $\chi$$_{n1}$ versus $H$ at $T$ = $T$$_{g}$. Red curve is the asymptotic fit to determine another critical exponent $\delta$. 
} \vspace{-4mm}
\end{figure}

First, we determine the nature of the spin glass transition in Tb$_{2}$Mo$_{2}$O$_{7}$. In a spin glass, the nonlinear part of dc susceptibilities are of special importance because of their sensitiveness to the freezing order parameter. Analysis of the nonlinear susceptibilities allows us to verify the thermodynamic (or equilibrium) nature of the spin glass transition. The nonlinear susceptibilities are written as the higher order terms in the following equations:\cite{Young,Martinez,Gingras}

\begin{eqnarray}
{M}/{H}({T})&=&{\chi_1}({T})-{\chi_3}({T}){H}^{2}+{O}({H}^{4})\\
                    &=&{\chi_1}({T})-{a}_{3}({T}){\chi_1}^{3}{H}^{2}+{O}({H}^{4})
\end{eqnarray}
\begin{eqnarray}
{\chi_n1}({T}, {H})&=& 1 - {M}({T}, {H})/{\chi_1}{H}
\end{eqnarray}

\noindent where $\chi$$_{1}$(T) is the linear susceptibility at temperature $T$, $\chi$$_{3}$(T) is the nonlinear susceptibility, coefficient $a$$_{3}$ = $\chi$$_{3}$/($\chi$$_{1}$)$^{3}$ and $\chi_n1$ is the net nonlinear susceptibility. Magnetization data on single crystal Tb$_{2}$Mo$_{2}$O$_{7}$ were obtained under conventional field-cooling condition in the field range of 10-10$^{4}$ Oe using a commercial magnetometer. The magnetic field was never decreased during the measurement and the sample was slowly cooled from 70 K to 10 K at a rate of 0.01 K/min. We followed the measurement procedure described in Ref. [5] of slow cooling under constant field, as it reduces the possible artifacts in the measurements due to magnetic hysteresis in the superconducting magnet of the magnetometer. It also allows a direct comparison of the static properties between Tb$_{2}$Mo$_{2}$O$_{7}$ and the isostructural spin glass Y$_{2}$Mo$_{2}$O$_{7}$. $\chi$$_{1}$($T$) at different temperatures were determined by fitting the M versus H curves at low fields. Due to the slow freezing mechanism of spin glass compared to the experimental observation time, it was not viable to analyze more than 2nd order term in the magnetization data. Therefore, equation (1) reduces to $\chi_3$($T$, $H$) $H$$^{2}$= 1 - $M(T, H)$/$\chi_1$$H$. Hence, $\chi_n1$($T$, $H$) becomes $\chi_3$($T$, $H$)$H$$^{2}$. In Fig. 1a, we have plotted the net nonlinear susceptibilities, $\chi_n1$($T$, $H$), at various temperatures as a function of $H^2$. Since we are interested in understanding the thermodynamic behavior near the spin freezing transition, therefore only experimental data from 40 K to $\simeq$ 24 K are shown in this figure. The maximum value of the field up to which $\chi$$_{n1}$(T) is linear in $H^2$, decreases rapidly as $T$ approaches $T_g$. It possibly arises due to the higher order corrections in the net susceptibility.\cite{Monod}  The linear portion of $\chi$$_{n1}$(T) at different temperatures were fitted with Eq. (2) to extract the coefficient $a$$_{3}$(T).

\begin{figure}
\centering
\includegraphics[width=9cm]{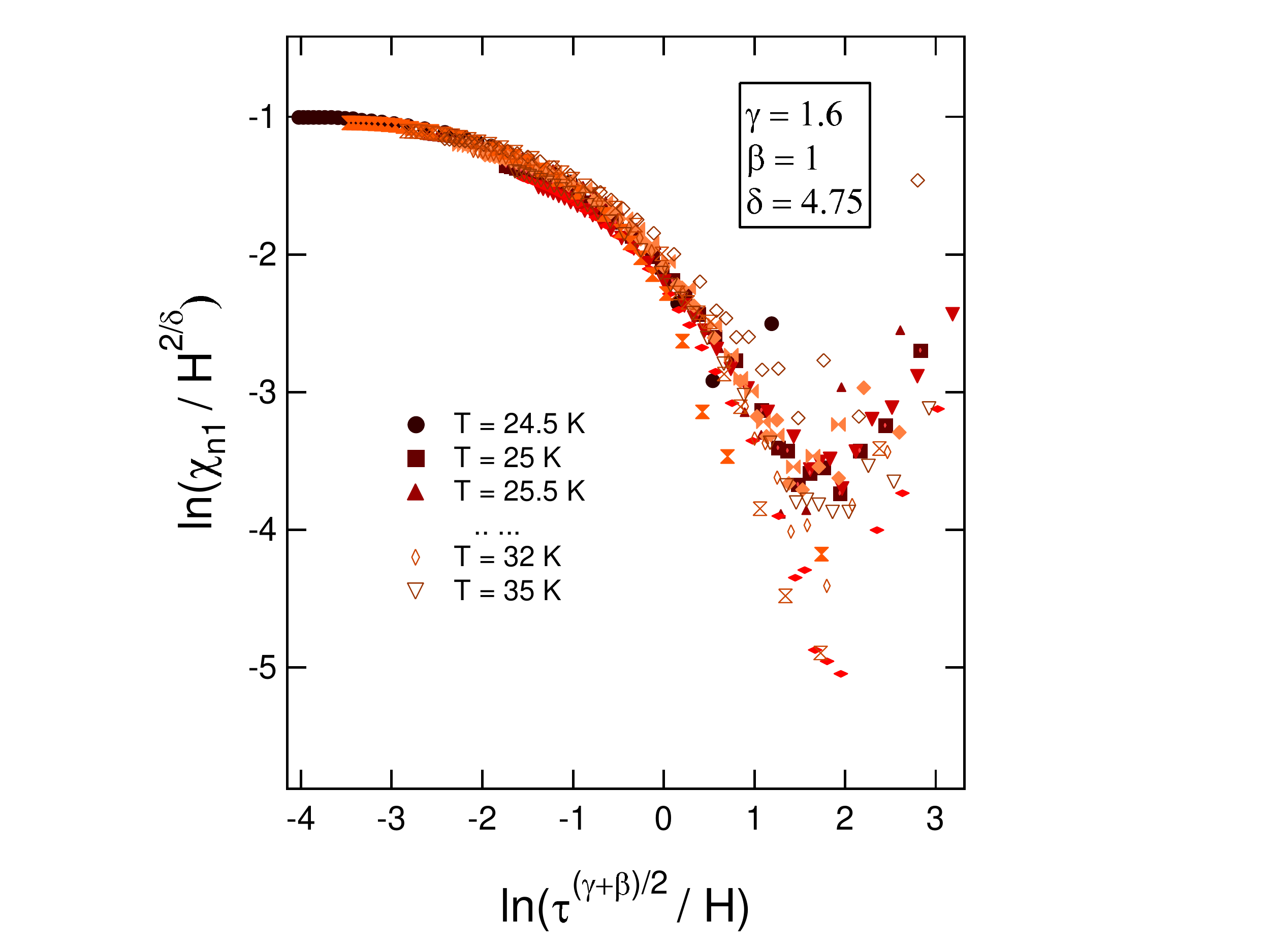} \vspace{-4mm}
\caption{(color online) Scaling behavior of net non-linear susceptibilities for a set of critical exponents: $\gamma$ = 1.6, $\beta$ = 1 and $\delta$ = 4.75.
} \vspace{-4mm}
\end{figure}

In a conventional spin glass, where the transition is an equilibrium phenomenon, the nonlinear susceptibilities exhibit a scaling behavior according to the single parameter, given by:\cite{Young,Martinez,Gingras}

\begin{eqnarray}
{\chi}_{n1} ({T}, {H})&=&{H}^{2/{\delta}} {f}({{\tau}^{({\gamma}+{\beta})/2}/H})
\end{eqnarray}

\noindent where $\tau$ = ($T$/$T_g$)- 1, $\gamma$ is the static critical exponent describing the divergent nature of magnetic susceptibility as a function of temperature and $\beta$ is the spin glass order parameter critical exponent. Determination of these critical exponents depends on the asymptotic nature of the arbitrary scaling function $f(x)$, with the boundary conditions $f(x)$ = $Constant$ as $x$$\rightarrow$~0 and $f(x)$ = $x$$^{-2\gamma/(\gamma+\beta)}$ as $x$$\rightarrow$~$\infty$. The nonlinear susceptibility, $\chi_n1$($T$, $H$), is expected to follow power-law dependence in both $T$ and $H$ with two independent static critical exponents $\gamma$ and $\delta$, respectively. The power law dependencies are described by the following expressions:\cite{Young,Martinez},

\begin{eqnarray}
{\chi}_{n1}(T)&{\propto}&{\tau}^{-\gamma}\\
{\chi}_{n1}(T={T_g, H})&{\propto}&{H}^{2/\delta}
\end{eqnarray}

\noindent Typically, the nonlinear susceptibilities diverge as $T$$\rightarrow$$T_g$ in a conventional spin glass. The two independent exponents, $\gamma$ and $\delta$, are related to the spin glass order parameter critical exponent $\beta$ via the following scaling relation:

\begin{eqnarray}
{\delta}&=& 1+({\gamma}/{\beta})
\end{eqnarray}

\noindent The above scaling relation represents a subtle test, arguably, of the true equilibrium phase transition in a spin glass system. 

We use the above formalism to explore the static nature of the spin glass transition in Tb$_{2}$Mo$_{2}$O$_{7}$. In Fig. 1b, we plot $a_3$ vs. $\tau$ for few different choices of glass transition temperatures $T_g$ $\in$ [22, 25] K. For the fitting purpose of Eq.(5), a fixed number of data points on each curve in the divergence regime were selected. As we see in Fig. 1c, $\gamma$ is found to vary in the range of [1.1, 2.4]. The best fit is obtained for $T_g$ = 24 K with the corresponding value of $\gamma$ = 1.6. While the estimated $T$$_{g}$ is consistent with the previous dc susceptibility measurements,\cite{Deepak,Gaulin}, the static critical exponent $\gamma$ is relatively smaller compared to the value ($\gamma$ $\simeq$ 2.25) in a conventional spin glass, which exhibits truely thermodynamic phase transition.\cite{Gingras,Bouchiat, Young}

Next, we determine another critical exponent $\delta$ by plotting ln($\chi$$_{n1}$, $T$ = $T_g$) versus ln($H$) in Fig. 1d. The best fit of the data using asymptotic expression (6) gives $\delta$= 3.67. If the spin glass transition in Tb$_{2}$Mo$_{2}$O$_{7}$ is indeed a true equilibrium phase transition at $T_g$ = 24 K, the nonlinear susceptibilities should follow the critical scaling behavior described by Eq.(4). Keeping the value of $\gamma$ constant, we vary other critical exponents $\delta$ and $\beta$ systematically to explore the scaling behavior. As shown in Fig.2, the nonlinear susceptibilities at different temperatures indeed exhibit the scaling collapse on one curve for a set of static critical exponents: $\gamma$ = 1.6, $\delta$ = 4.75 and $\beta$ = 1. At large $x$ values, correspond to higher temperatures and smaller fields, some data scatter from the scaling curve due to the large errors associated with the smaller nonlinear susceptibilities. Although an asymptotic behavior is observed in the scaling plot of Fig. 2, but the scaling relation in Eq. (7) is not fulfilled with that set of critical exponents. It suggests the absence of static critical behavior in this compound. This fact, together with the weak divergence of $a_3$ coefficient as $T_g$ is approached from above, indicates an unconventional spin glass transition in Tb$_{2}$Mo$_{2}$O$_{7}$. The discrepancies in the static critical exponents (between the estimated and the scaling values)  leading to the unconventional spin glass behavior can be attributed, arguably, to the formation of small ferromagnetic clusters with short-range order, which ultimately enhances the $\chi$$_{n1}$ considerably and led to strong but non-critical background temperature dependence. Similar behavior has been observed in some canonical spin glass systems, which exibits non-equilibrium transition.\cite{Binder} We would like to point out that the spin glass state in Tb$_{2}$Mo$_{2}$O$_{7}$ is clearly different from a regular ferromagnet with disorder, as no chemical disorder is found in this compound. Also, the small amount of disorder to an ordered system (a regular ferromagnet) should not affect the phase transition and the associated static critical behavior.\cite{Kaul} 

\begin{figure}
\centering
\includegraphics[width=10cm]{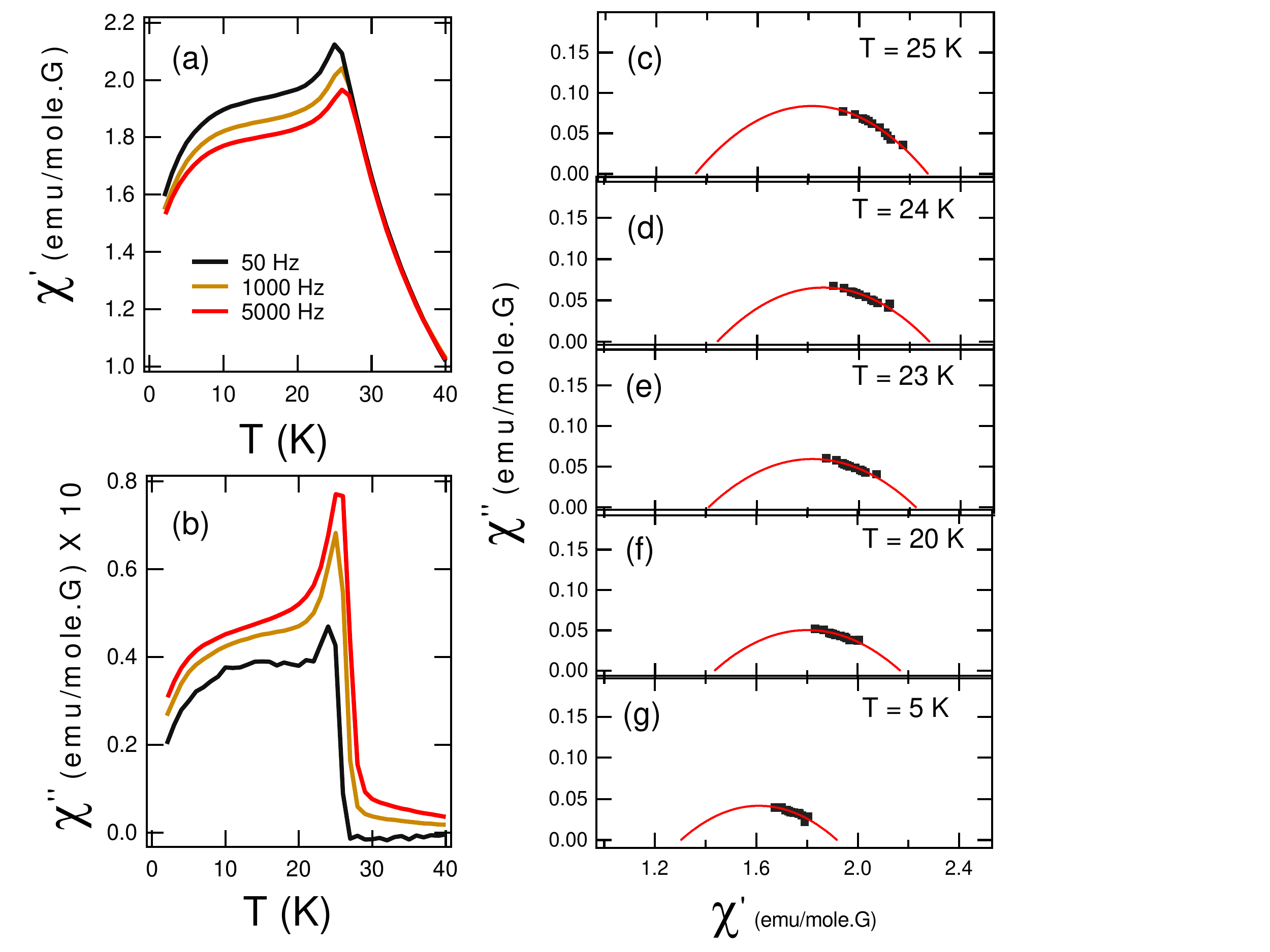} \vspace{-4mm}
\caption{(color online) (a) and (b) Real and imaginary parts of ac-susceptibilities at few charecteristic frequencies. Sharp cusps, indicating the freezing transition, can clearly be seen at all measured frequencies in fig. (a). (c-g) Cole-cole diagrams at few different temperatures across $T_g$. Solid lines are fit to Eq. 9.
} \vspace{-4mm}
\end{figure}

The spin glass transition in short range ordered ferromagnetic clusters are often accompanied by a statistical distribution of relaxation times, as each cluster acts as an independent unit.\cite{Huser} The distribution of spin relaxation times can be extracted from the analysis of complex (ac) susceptibilities. We have performed the complex susceptibility measurements on Tb$_{2}$Mo$_{2}$O$_{7}$ using a superconducting quantum interference device in the frequency range of 10 - 10$^{4}$ Hz. The real and imaginary parts of $\chi$ at few charectarisitc frequencies are plotted in Fig. 3a-b. The $\chi^{'}$(T) curves show typical cusps whose maxima shift to higher temperatures with increasing frequencies, as expected for a spin glass.\cite{Mydosh} In addition, $\chi^{''}$ clearly shows a steep rise above zero below $T$ $\leq$ 26 K, indicating the existence of relaxation process in this system. The distribution of spin relaxation times in magnetic clusters is extracted by plotting the real and imaginary parts of complex susceptibilities in an Argand diagram, as described by the Cole-Cole formalism.\cite{Cole,Dekker} According to that, the complex susceptibility can be phenomenologically expressed as:

\begin{eqnarray}
{\chi}&=&{\chi_s}+ \frac{{\chi_0}-{\chi_s}}{1+(i{\omega}{\tau_c})^{1-{\alpha}}}
\end{eqnarray}

\noindent where $\chi_0$ and $\chi_s$ are the isothermal ($\omega$ = 0) and adiabatic ($\omega$$\rightarrow$0) susceptibilities, respectively, and $\tau_c$ is the median relaxation time around which a distribution of relaxation times is assumed. The parameter $\alpha$ determines the width of the distribution, such that $\alpha$ = 1 corresponds to an infinitely wide distribution and $\alpha$ = 0 returns the Debye equation of single relaxation time. In the Argand diagram, a single relaxation time is manifested by a perfectly symmetrical full half-circle, whereas a distribution of relaxation times in magnetic clusters at temperature $T$ is usually exhibited by a flattened semi-circular arc.\cite{Dekker} Equation (8) can be further decomposed into $\chi$$^{'}$ and $\chi$$^{''}$ to obtain the relations:\cite{Mydosh}

\begin{eqnarray*}
{\chi}^{''}=- \frac{{\chi_0}-{\chi_s}}{2tan[(1-{\alpha}){\pi}/2]}\\
                \\+\surd{[({\chi_0}-{\chi}^{'})({\chi}^{'}-{\chi_s})+\frac{({\chi_0}-{\chi_s})^{2}}{4tan^{2}((1-{\alpha}){\pi}/2)}]}
\end{eqnarray*}
\begin{eqnarray}
{\chi}^{''}=- \frac{{\chi_0}-{\chi_s}}{2}\frac{cos({\pi}{\alpha}/2)}{cosh[(1-{\alpha})ln({\omega}{\tau_c})]+sin({\pi}{\alpha}/2)}
\end{eqnarray}

Argand diagrams of $\chi^{''}$ vs $\chi^{'}$ at various temperatures are illustrated in Fig. 3c-g. We use equation 9a to fit the Argand diagrams in Fig. 3c-g with three adjustable parameters, $\chi_0$, $\chi_s$ and $\alpha$. The maxima of the diagrams give $\omega$$\tau_c$ = 1, while the flatness of the arc is a measure of the width of the distribution of relaxation times. The data points at different frequencies, in these figures, fall on semi-circular arcs of varying diameters, that tend to flatten as the temperature is reduced. This behavior indicates a distribution of relaxation times in this system. However, unlike some other spin glass systems, the data points at different frequnecies do not shift significantly across the maxima in $\chi$$^{''}$ as the system passes through the spin glass transition. This indicates an unusual spin dynamics feature in Tb$_{2}$Mo$_{2}$O$_{7}$, which qualitatively suggests the insufficient freezing of spins well below the glass transition temperature. We determine the actual value of $\tau_c$ at each temperature by separately fitting $\chi$$^{''}$ vs ln($\nu$), as shown in Fig. 4a, using the refined values of $\chi_0$, $\chi_s$ and $\alpha$ in Eq. 9b. In the Cole-Cole plots of Fig. 3c-g, the shape of the arc changes as the temperature is reduced. It clearly indicates that the distribution of relaxation times in magnetic clusters changes with temperature. Similar behavior is observed in the $\chi$$^{''}$ vs ln($\nu$) plots, where we see increase in the width of the curves in the vicinity of $T$$_{g}$ = 24 K. The quantitative measure of the median relaxation time $\tau_c$ and its distribution representative, $\alpha$, are presented in Fig. 4b as a function of temperature. Both ln($\tau_c$) and $\alpha$ increase as the temperature is reduced. At $T$$\leq$ $T$$_{g}$, they saturate around $\simeq$ -6 and 0.55, respectively. These analysis indicate that the system is not sufficiently frozen below $T_g$ and significant spin dynamics is still present. It also explains why the spin glass transtiion in Tb$_{2}$Mo$_{2}$O$_{7}$ is a non-equilibrium phenomenon, as depicted earlier by the non-compliance of scaling law by static critical exponents deduced from the nonlinear susceptibilities.

\begin{figure}
\centering
\includegraphics[width=8cm]{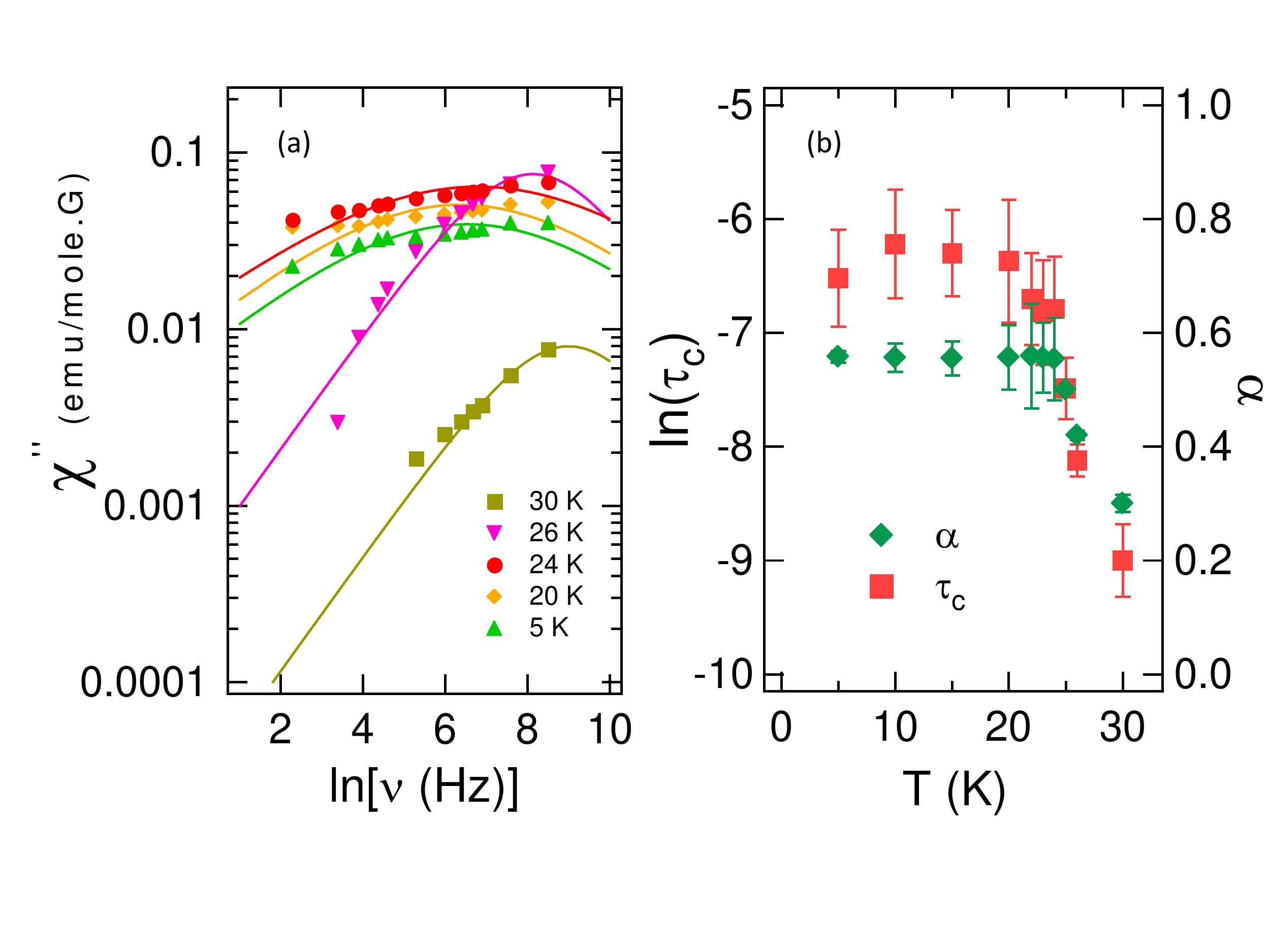} \vspace{-4mm}
\caption{(color online) (a) $\chi$$^{''}$ as a function of frequency at several temperatures. The solid lines are fit to the data using Eq. 9 (see text). (b) Temperature dependence of the median relaxation time $\tau$$_{c}$ and parameter $\alpha$, reflecting the probability distribution of relaxation times.
} \vspace{-4mm}
\end{figure}

In general, a spin glass state arises in a material where the combination of randomness and frustration prevents the development of long range magnetic order. However, in some materials, such as Tb$_{2}$Mo$_{2}$O$_{7}$ or, Y$_{2}$Mo$_{2}$O$_{7}$ or, Gd$_{3}$Ga$_{5}$O$_{12}$, no evidence of chemical disorder is found. We compare our results with some of these unconventional as well as conventional or random spin glasses. The power law divergence coefficient, $\gamma$, of nonlinear susceptibility is found to vary in the range of [2.0, 4.0] in a spin glass compound. Many of them exhibit thermodynamic transition to the spin glass state; random spin glasses and Y$_{2}$Mo$_{2}$O$_{7}$ are notable examples.\cite{Young,Bouchiat,Gingras} However, GGG or SCGO do not exhibit the equilibrium spin glass behavior.\cite{Schiffer,Martinez} In these materials, the divergence of $\chi$$_{n1}$ or, the scaling behavior were not found to be physically meaningful. The random spin glasses also tends to exhibit complete freezing (i.e. $\alpha$$\rightarrow$1) as T$\rightarrow$0 K. Our investigation of the spin glass properties in Tb$_{2}$Mo$_{2}$O$_{7}$ reveals the nonconventional nature of transition, as manifested by a smaller $\gamma$ = 1.6, compared to both conventional and some of the unconventional spin glass systems, a temperature dependent distribution of relaxation times in magnetic clusters and finite spin dynamics below the glass transition. Our results are also consistent with a previous $\mu$SR measurement\cite{Dunsiger}, where significantly smaller relaxation time in the spin glass state, indicating the active presence of spin dynamics, was reported. This behavior is in strong contrast to the conventional spin glass transition in Y$_{2}$Mo$_{2}$O$_{7}$, where the magnetic properties are dominated by the antiferromagnetically coupled Mo sublattice only.\cite{Jason} Due to the small Mo$^{4+}$ moment compared to Tb$^{3+}$ moment, the Tb-Tb correlation is the predominant contributor in the magnetic measurements. Recent reports on the study of various possible interactions in this compound suggest that the competition of FM/AFM interactions (between Tb-Tb and Tb-Mo correlations or, Tb-Tb and Mo-Mo correlations) could be playing a crucial role in the formation of dynamic spin glass state in Tb$_{2}$Mo$_{2}$O$_{7}$.\cite{Ehler,Jiang} Our results can be useful in understanding the anomalous spin glass transition in other geometrically frustrated pyrochlores with short-range ferromagnetic order, such as Tb$_{2}$Sn$_{2}$O$_{7}$\cite{Cava} or Y$_{2}$Mn$_{2}$O$_{7}$\cite{Reimers}. The work at MIT was supported by the Department of Energy (DOE) under Grant No. DE-FG02-07ER46134. D. K. S. acknowledges support form NSF under agreement no. DMR-0944772.

\bibliography{TMO2_Revised}

\begin{thebibliography}{26}
\expandafter\ifx\csname natexlab\endcsname\relax\def\natexlab#1{#1}\fi
\expandafter\ifx\csname bibnamefont\endcsname\relax
  \def\bibnamefont#1{#1}\fi
\expandafter\ifx\csname bibfnamefont\endcsname\relax
  \def\bibfnamefont#1{#1}\fi
\expandafter\ifx\csname citenamefont\endcsname\relax
  \def\citenamefont#1{#1}\fi
\expandafter\ifx\csname url\endcsname\relax
  \def\url#1{\texttt{#1}}\fi
\expandafter\ifx\csname urlprefix\endcsname\relax\def\urlprefix{URL }\fi
\providecommand{\bibinfo}[2]{#2}
\providecommand{\eprint}[2][]{\url{#2}}

\bibitem[{\citenamefont{Helton\emph{, et al.}}(2007)}]{Joel}
\bibinfo{author}{\bibfnamefont{J.~S.} \bibnamefont{Helton\emph{, et al.}}},
  \bibinfo{journal}{Phys. Rev. Lett.} \textbf{\bibinfo{volume}{98}},
  \bibinfo{pages}{107204} (\bibinfo{year}{2007}).

\bibitem[{\citenamefont{Ramirez et~al.}(1999)\citenamefont{Ramirez, Hayashi,
  Cava, Siddharthan, and Shastry}}]{Ramirez2}
\bibinfo{author}{\bibfnamefont{A.~P.} \bibnamefont{Ramirez}},
  \bibinfo{author}{\bibfnamefont{A.}~\bibnamefont{Hayashi}},
  \bibinfo{author}{\bibfnamefont{R.~J.} \bibnamefont{Cava}},
  \bibinfo{author}{\bibfnamefont{R.}~\bibnamefont{Siddharthan}},
  \bibnamefont{and} \bibinfo{author}{\bibfnamefont{B.~S.}
  \bibnamefont{Shastry}}, \bibinfo{journal}{Nature}
  \textbf{\bibinfo{volume}{399}}, \bibinfo{pages}{333} (\bibinfo{year}{1999}).

\bibitem[{\citenamefont{Gardner et~al.}(1999)\citenamefont{Gardner, Gaulin,
  Lee, Broholm, Raju, and Greedan}}]{Jason2}
\bibinfo{author}{\bibfnamefont{J.~S.} \bibnamefont{Gardner}},
  \bibinfo{author}{\bibfnamefont{B.~D.} \bibnamefont{Gaulin}},
  \bibinfo{author}{\bibfnamefont{S.~H.} \bibnamefont{Lee}},
  \bibinfo{author}{\bibfnamefont{C.}~\bibnamefont{Broholm}},
  \bibinfo{author}{\bibfnamefont{N.~P.} \bibnamefont{Raju}}, \bibnamefont{and}
  \bibinfo{author}{\bibfnamefont{J.~E.} \bibnamefont{Greedan}},
  \bibinfo{journal}{Phys Rev Lett} \textbf{\bibinfo{volume}{83}},
  \bibinfo{pages}{211} (\bibinfo{year}{1999}).

\bibitem[{\citenamefont{Gaulin et~al.}(1992)\citenamefont{Gaulin, Reimers,
  Mason, Greedan, and Tun}}]{Gaulin}
\bibinfo{author}{\bibfnamefont{B.~D.} \bibnamefont{Gaulin}},
  \bibinfo{author}{\bibfnamefont{J.~N.} \bibnamefont{Reimers}},
  \bibinfo{author}{\bibfnamefont{T.~E.} \bibnamefont{Mason}},
  \bibinfo{author}{\bibfnamefont{J.~E.} \bibnamefont{Greedan}},
  \bibnamefont{and} \bibinfo{author}{\bibfnamefont{Z.}~\bibnamefont{Tun}},
  \bibinfo{journal}{Phys Rev Lett} \textbf{\bibinfo{volume}{69}},
  \bibinfo{pages}{3244} (\bibinfo{year}{1992}).

\bibitem[{\citenamefont{Gingras\emph{, et al.}}(1997)}]{Gingras}
\bibinfo{author}{\bibfnamefont{M.~J.~P.} \bibnamefont{Gingras\emph{, et al.}}},
  \bibinfo{journal}{Phys Rev Lett} \textbf{\bibinfo{volume}{78}},
  \bibinfo{pages}{947} (\bibinfo{year}{1997}).

\bibitem[{\citenamefont{Schiffer\emph{, et al.}}(1995)}]{Schiffer}
\bibinfo{author}{\bibfnamefont{P.}~\bibnamefont{Schiffer\emph{, et al.}}},
  \bibinfo{journal}{Phys Rev Lett} \textbf{\bibinfo{volume}{74}},
  \bibinfo{pages}{2379} (\bibinfo{year}{1995}).

\bibitem[{\citenamefont{Ramirez\emph{, et al.}}(1990)}]{Ramirez}
\bibinfo{author}{\bibfnamefont{A.~P.} \bibnamefont{Ramirez\emph{, et al.}}},
  \bibinfo{journal}{Phys Rev Lett} \textbf{\bibinfo{volume}{64}},
  \bibinfo{pages}{2070} (\bibinfo{year}{1990}).

\bibitem[{\citenamefont{Binder and Young}(1986)}]{Young}
\bibinfo{author}{\bibfnamefont{K.}~\bibnamefont{Binder}} \bibnamefont{and}
  \bibinfo{author}{\bibfnamefont{A.~P.} \bibnamefont{Young}},
  \bibinfo{journal}{Rev. Mod. Phys.} \textbf{\bibinfo{volume}{58}},
  \bibinfo{pages}{801} (\bibinfo{year}{1986}).

\bibitem[{\citenamefont{Andreanov\emph{, et al.}}(2010)}]{Andreanov}
\bibinfo{author}{\bibfnamefont{A.}~\bibnamefont{Andreanov\emph{, et al.}}},
  \bibinfo{journal}{Phys. Rev. B} \textbf{\bibinfo{volume}{81}},
  \bibinfo{pages}{014406} (\bibinfo{year}{2010}).

\bibitem[{\citenamefont{Martinez\emph{, et al.}}(1994)}]{Martinez}
\bibinfo{author}{\bibfnamefont{B.}~\bibnamefont{Martinez\emph{, et al.}}},
  \bibinfo{journal}{Phys. Rev. B} \textbf{\bibinfo{volume}{50}},
  \bibinfo{pages}{15779} (\bibinfo{year}{1994}).

\bibitem[{\citenamefont{Singh\emph{, et al.}}(2008)}]{Deepak}
\bibinfo{author}{\bibfnamefont{D.~K.} \bibnamefont{Singh\emph{, et al.}}},
  \bibinfo{journal}{Phys. Rev. B} \textbf{\bibinfo{volume}{78}},
  \bibinfo{pages}{220405(R)} (\bibinfo{year}{2008}).

\bibitem[{\citenamefont{Ehlers\emph{, et al.}}(2010)}]{Ehler}
\bibinfo{author}{\bibfnamefont{G.}~\bibnamefont{Ehlers\emph{, et al.}}},
  \bibinfo{journal}{Phys. Rev. B} \textbf{\bibinfo{volume}{81}},
  \bibinfo{pages}{224405} (\bibinfo{year}{2010}).

\bibitem[{\citenamefont{Jiang\emph{, et al.}}(2011)}]{Jiang}
\bibinfo{author}{\bibfnamefont{Y.}~\bibnamefont{Jiang\emph{, et al.}}},
  \bibinfo{journal}{J. Phys: Condens. Matter} \textbf{\bibinfo{volume}{23}},
  \bibinfo{pages}{164214} (\bibinfo{year}{2011}).

\bibitem[{\citenamefont{Greedan et~al.}(1991)\citenamefont{Greedan, Reimers,
  Stager, and Penny}}]{Greedan}
\bibinfo{author}{\bibfnamefont{J.~E.} \bibnamefont{Greedan}},
  \bibinfo{author}{\bibfnamefont{J.~N.} \bibnamefont{Reimers}},
  \bibinfo{author}{\bibfnamefont{C.~V.} \bibnamefont{Stager}},
  \bibnamefont{and} \bibinfo{author}{\bibfnamefont{S.~L.} \bibnamefont{Penny}},
  \bibinfo{journal}{Phys. Rev. B} \textbf{\bibinfo{volume}{43}},
  \bibinfo{pages}{5682} (\bibinfo{year}{1991}).

\bibitem[{\citenamefont{Monod\emph{, et al.}}(1982)}]{Monod}
\bibinfo{author}{\bibfnamefont{P.}~\bibnamefont{Monod\emph{, et al.}}},
  \bibinfo{journal}{J. Phys. (Paris) Lett.} \textbf{\bibinfo{volume}{43}},
  \bibinfo{pages}{145} (\bibinfo{year}{1982}).

\bibitem[{\citenamefont{Bouchiat}(1986)}]{Bouchiat}
\bibinfo{author}{\bibfnamefont{H.}~\bibnamefont{Bouchiat}},
  \bibinfo{journal}{J. Phys. (Paris) Lett.} \textbf{\bibinfo{volume}{47}},
  \bibinfo{pages}{71} (\bibinfo{year}{1986}).

\bibitem[{\citenamefont{Binder}(1982)}]{Binder}
\bibinfo{author}{\bibfnamefont{K.}~\bibnamefont{Binder}}, \bibinfo{journal}{Z.
  Phys. B} \textbf{\bibinfo{volume}{48}}, \bibinfo{pages}{319}
  (\bibinfo{year}{1982}).

\bibitem[{\citenamefont{Kaul}(1985)}]{Kaul}
\bibinfo{author}{\bibfnamefont{S.~N.} \bibnamefont{Kaul}}, \bibinfo{journal}{J.
  Mag. Mag. Mat.} \textbf{\bibinfo{volume}{53}}, \bibinfo{pages}{5}
  (\bibinfo{year}{1985}).

\bibitem[{\citenamefont{Huser\emph{, et al.}}(1986)}]{Huser}
\bibinfo{author}{\bibfnamefont{D.}~\bibnamefont{Huser\emph{, et al.}}},
  \bibinfo{journal}{J. Phys. C: Solid State Phys.}
  \textbf{\bibinfo{volume}{19}}, \bibinfo{pages}{3697} (\bibinfo{year}{1986}).

\bibitem[{Myd()}]{Mydosh}
\bibinfo{note}{A. Mydosh "Spin Glasses: An Experimental Introduction", CRC 1st
  edition (1993)}.

\bibitem[{\citenamefont{Cole and Cole}(1941)}]{Cole}
\bibinfo{author}{\bibfnamefont{K.~S.} \bibnamefont{Cole}} \bibnamefont{and}
  \bibinfo{author}{\bibfnamefont{R.~H.} \bibnamefont{Cole}},
  \bibinfo{journal}{J. Chem. Phys.} \textbf{\bibinfo{volume}{9}},
  \bibinfo{pages}{341} (\bibinfo{year}{1941}).

\bibitem[{\citenamefont{Dekker\emph{, et al.}}(1989)}]{Dekker}
\bibinfo{author}{\bibfnamefont{C.}~\bibnamefont{Dekker\emph{, et al.}}},
  \bibinfo{journal}{Phys. Rev. B} \textbf{\bibinfo{volume}{40}},
  \bibinfo{pages}{11243} (\bibinfo{year}{1989}).

\bibitem[{\citenamefont{Dunsiger\emph{, et al.}}(1996)}]{Dunsiger}
\bibinfo{author}{\bibfnamefont{S.~R.} \bibnamefont{Dunsiger\emph{, et al.}}},
  \bibinfo{journal}{Phys Rev B} \textbf{\bibinfo{volume}{54}},
  \bibinfo{pages}{9019} (\bibinfo{year}{1996}).

\bibitem[{\citenamefont{Gardner et~al.}(2010)\citenamefont{Gardner, Gingras,
  and Greedan}}]{Jason}
\bibinfo{author}{\bibfnamefont{J.}~\bibnamefont{Gardner}},
  \bibinfo{author}{\bibfnamefont{M.}~\bibnamefont{Gingras}}, \bibnamefont{and}
  \bibinfo{author}{\bibfnamefont{J.}~\bibnamefont{Greedan}},
  \bibinfo{journal}{Rev. Mod. Phys.} \textbf{\bibinfo{volume}{82}},
  \bibinfo{pages}{53} (\bibinfo{year}{2010}).

\bibitem[{\citenamefont{Dahlberg\emph{, et al.}}(2011)}]{Cava}
\bibinfo{author}{\bibfnamefont{M.~L.} \bibnamefont{Dahlberg\emph{, et al.}}},
  \bibinfo{journal}{Phys Rev B} \textbf{\bibinfo{volume}{83}},
  \bibinfo{pages}{140410(R)} (\bibinfo{year}{2011}).

\bibitem[{\citenamefont{Reimers et~al.}(1991)\citenamefont{Reimers, Greedan,
  Kremer, Gmelin, and Subramanian}}]{Reimers}
\bibinfo{author}{\bibfnamefont{J.~N.} \bibnamefont{Reimers}},
  \bibinfo{author}{\bibfnamefont{J.~E.} \bibnamefont{Greedan}},
  \bibinfo{author}{\bibfnamefont{R.~K.} \bibnamefont{Kremer}},
  \bibinfo{author}{\bibfnamefont{E.}~\bibnamefont{Gmelin}}, \bibnamefont{and}
  \bibinfo{author}{\bibfnamefont{M.~A.} \bibnamefont{Subramanian}},
  \bibinfo{journal}{Phys. Rev. B} \textbf{\bibinfo{volume}{43}},
  \bibinfo{pages}{3387} (\bibinfo{year}{1991}).

\end{thebibliography}
\end{document}